\documentclass{ws-ijmpcs}

\begin{document}

\markboth{S. P. Kim \& W. Kim}
{Will quantum cosmology resurrect chaotic inflation model?}

%
\catchline{}{}{}{}{}
%

\title{WILL QUANTUM COSMOLOGY RESURRECT CHAOTIC INFLATION MODEL?}

\author{SANG PYO KIM${}^*$ and WON KIM}

\address{Department of Physics, Kunsan National University, Kunsan 54150, Korea\\
${}^*$sangkim@kunsan.ac.kr}

\maketitle

\begin{history}
\received{\today}
\revised{Day Month Year}
\published{Day Month Year}
\end{history}

\begin{abstract}
The single field chaotic inflation model with a monomial power greater than one seems to be ruled out by the recent Planck and WMAP CMB data while Starobinsky model with a higher curvature term seems to be a viable model. Higher curvature terms being originated from quantum fluctuations, we revisit the quantum cosmology of the Wheeler-DeWitt equation for the chaotic inflation model. The semiclassical cosmology emerges from quantum cosmology with fluctuations of spacetimes and matter when the wave function is peaked around the semiclassical trajectory with quantum corrections a la the de Broglie-Bohm pilot theory.
\end{abstract}

\ccode{PACS numbers:}

\section{Introduction}	

The first direct detection of gravitational waves from the merger of binary black holes,\cite{LIGO col} the most important scientific discovery of the 21st century, will open a new window of gravitational waves probe to explore the early history of the universe far before the last scattering epoch and hopefully around the origin of the universe. The tensorial part of spacetime fluctuations gives rise to gravitational waves, whose observation will reconstruct the spacetime and its evolution history. If the universe would have evolved back in time, the singularity theorem implies the Big Bang, which belongs to a quantum gravity regime.\cite{hawking-penrose} The inflationary spacetime, the most viable cosmological model, would have the singularity.\cite{BGV} The singularity theorem may raise a few fundamental questions in general relativity: what is the quantum spacetime and geometry including the Big Bang? how to quantize the spacetime as well as matter fields, that is, what is quantum gravity and quantum cosmology? how do a classical universe and the unitary quantum field theory emerge from quantum cosmology?

\begin{figure}[pb]
\centerline{\includegraphics[width=7.0cm]{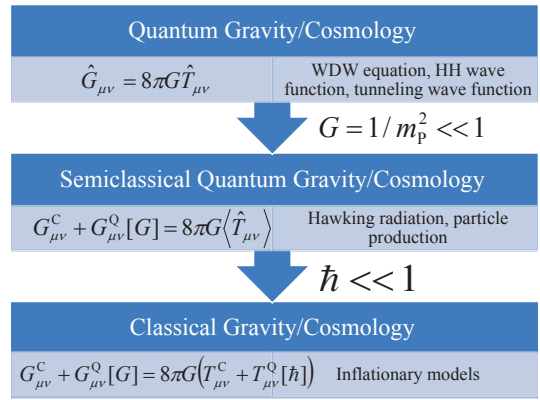}}
\vspace*{8pt}
\caption{Emergence of classical gravity from quantum gravity and quantum effects. \label{QSC}}
\end{figure}

Quantum cosmology is a quantum gravity theory for the universe, which quantizes simultaneously the spacetime geometry and matter fields. Two typical approaches to quantum cosmology are quantum geometrodynamics based on the Wheeler-DeWitt (WDW) equation\cite{dewitt67} and the no-boundary (HH) wave function, a path integral over spacetime manifolds and matters, by Hartle and Hawking.\cite{hartle-hawking,hawking84} In this proceedings we will not consider loop quantum gravity and other quantum gravity models (for review and references, see Ref.~\refcite{coule}). In quantum geometrodynamics the WDW equation is the relativistic field equation in a superspace of the spacetime geometry and the matter fields, in which both diffeomorphic invariant spacetime variables and matter fields are quantized. The quantum geometrodynamics has the advantage in predicting quantum gravity effects that can be tested by the current observational data based on classical cosmology such as CMB etc since as summarized in Fig. \ref{QSC}, the standard cosmology can be derived from the semiclassical quantum cosmology, which in turn can be derived from the WDW equation. In each stage for transitions from quantum to semiclassical and then to classical gravity, spacetime and matter fluctuations leave imprints of quantum gravity effects, which differ from the quantum field theory in the curved spacetime.

On the other hand, the HH no-boundary wave function sums over all compact four-dimensional Euclidean geometry and matter fields with a three-dimensional boundary to a Lorenztian geometry.\cite{hartle-hawking,hawking84} The HH wave function has the advantage of incorporating a boundary condition (initial condition) not to mention the quantum law. In fact, the path integral is peaked around the WDW equation at the tree level. From the view point of the standard cosmology, Page summarized the predictions of the HH wave function:\cite{page06} inflation of the universe to large size,\cite{hawking84} prediction of the near-critical density,\cite{hawking84,hawking-page} inhomogeneities starting in ground states,\cite{halliwell-hawking} arrow of time and initial low entropy,\cite{hawking85,page85,HLL,kim-kim94,kim-kim95} and decoherence and classicality of the universe.\cite{kiefer87,kim92} Starobinsky argued that inflation scenario relates quantum gravity and quantum cosmology to astronomical observations and produces (non-universal) arrow of time for our universe.\cite{starobinsky}

In this paper, we review the quantum cosmology of the Friedmann-Robertson-Walker (FRW) universe minimally coupled to a massive scalar field and argue that the quantum gravity effects may resurrect the chaotic inflation model with a massive scalar model. The recent Planck data rules out the single field chaotic model with power greater than one, including the massive scalar.\cite{Planck2013} The Starobinsky inflation model of $R+ \alpha R^2$, however, which leads to a de Sitter-type acceleration without inflaton, is the most favored by the Planck data.\cite{starobinsky80} As noted by Starobinsky, $R^2$ comes from spacetime fluctuations due to quantum matter. It has been noticed that $R^2$-term is equivalent to a scalar field under a conformal transformation, $\tilde{g}_{\mu \nu} = (1+ 2 \alpha R) g_{\mu \nu}$ and $\Psi = \sqrt{3/2} \ln (1+ 2 \alpha R)$.\cite{whitt84,maeda88}  Further, as shown in Fig. \ref{QSC}, the quantum gravity effects from quantum cosmology, which have both quantum corrections from spacetime as well as the expectation value of the energy-momentum stress tensor, differs from those from quantum field theory in a fixed curved spacetime. It is thus interesting and timely to revisit the quantum cosmology with a massive scalar field and to investigate the quantum gravity effects.

\section{Why Massive Scalar Quantum Cosmology?} \label{mass quan cos}

The FRW universe with a scale factor $a = e^{\alpha}$ and an inflaton $\phi$ has a superspace with the supermetric
\begin{eqnarray}
ds^2 = - da^2 + a^2 \phi^2 \label{sup met}
\end{eqnarray}
and leads to the super-Hamiltonian constraint
\begin{eqnarray}
H(a, \phi) = - \Bigl( \pi_a^2 + V_{\rm G} (a) \Bigr) + \frac{1}{a^2} \Bigl( \pi_{\phi}^2 + 2 a^6 V (\phi) \Bigr) = 0.
\end{eqnarray}
Then, the WDW equation takes the form (see, for instance, Ref. \refcite{kim92} and for a recent review and references see Refs. \refcite{kim13,kim14})
\begin{eqnarray}
\Bigl[ - \nabla^2 - V_{\rm G} +2 a^4 V(\phi) \Bigr] \Psi (a, \phi) = 0, \label{wdw eq}
\end{eqnarray}
where
\begin{eqnarray}
\nabla^2 = - a^{-1} \frac{\partial}{\partial a} \Bigl(a \frac{\partial}{\partial a} \Bigr) + \frac{1}{a^2} \frac{\partial^2}{\partial \phi^2}, \quad
V_{\rm G} (a) = k a^2 - 2 \Lambda a^4.
\end{eqnarray}
Note that the WDW equation becomes a relativistic wave equation in the superspace, which is generically true for any spacetime with more than two degrees of freedom or with a matter field.

To compare the predictions of quantum cosmology with the inflation scenario and current observational data, we may introduce perturbations of spacetime and/or matter. The Fourier-modes $f_{k}$ of $\phi$-fluctuations, for instance, have the wave function $\Psi (\alpha, \phi, f_{k})$.\cite{halliwell-hawking} We assume the $\phi$-derivatives to be much smaller than the $\alpha$-derivatives, which corresponds to a slow-roll approximation in the inflation scenario. In the geometry belonging to a classical regime, the wave function of the WDW equation becomes a wavepacket and is peaked around a semi-classical trajectory, along which we may apply the the Born-Oppenheimer interpretation.\cite{kim95,kim96,BFV,kim97} Recently, Kiefer and Kr\"{a}mer have found the power spectrum corrected by quantum cosmology\cite{kiefer-kramer,kk-essay}
\begin{eqnarray}
\Delta_{(1)}^2 (k) = \underbrace{ \Delta_{(0)}^2 (k)}_{\rm classical~cosmology} \underbrace{\Bigl(1 - \frac{43.56}{k^3} \frac{H^2}{m_P^2} \Bigr)^{-3/2} \Bigl(1 - \frac{189.18}{k^3} \frac{H^2}{m_P^2} \Bigr)}_{\rm correction~from~quantum~cosmology}, \label{kk sp}
\end{eqnarray}
where $\Delta_{(0)}^2(k)$ is the spectrum from classical theory. Note that the power spectrum (\ref{kk sp}) is suppressed at large scales and shows a weaker upper bound than the tensor-to-scalar ratio.

From the view point of density perturbations, there is a formulation of gauge invariant perturbations in quantum cosmology. Choosing gauge invariant perturbations is equivalent to selecting the diffeomorphism invariant variables for the superspace. The gauge invariant perturbations still have advantage in interpreting the observational data in cosmology. Recently, the gauge invariant super-Hamiltonian and super-momenta constraints have been introduced in terms of Mukhanov-Sasaki variables\cite{pinho-pinto,PSS,CFMO,CMM}. Then, the classical cosmology from the quantum cosmology may give a complete description of density perturbations with quantum effects included for CMB data.

\section{Second Quantized Universes} \label{quan univ}

The WDW equation (\ref{wdw eq}) is a relativistic wave in the superspace (\ref{sup met}), in which $a$ plays the role of an intrinsic time. As for quantum field theory in a curved spacetime, the WDW equation evolves an initial wave function  $\Psi (a_0, \phi)$ to a final one $\Psi (a, \phi)$. The Cauchy initial value problem of the WDW equation has been well elaborated.\cite{kim92,kim-page92} Note that the HH wave function has a different Cauchy surface, $\alpha = \pm \phi$.
Then, a question is how to prescribe the boundary condition that leads to the present universe. For the single-field inflation model with a monomial potential, Kim observed that the eigenfunctions for the inflaton\cite{kim92,kim-page92}
\begin{eqnarray}
H_{\rm M} (\phi, a) \Phi_n (\phi, a) = E_n (a) \Phi_n (\phi, a), \quad V(\phi) = \frac{\lambda_{2p}}{2p} \phi^{2p}
\end{eqnarray}
obey the Symanzik scaling-law
\begin{eqnarray}
E_n (a) = \Bigl(\frac{\lambda_{2p}a^6}{p} \Bigr)^{\frac{1}{p+1}} \epsilon_n, \quad \Phi_n (\phi, a)= \Bigl(\frac{\lambda_{2p}a^6}{p} \Bigr)^{\frac{1}{4(p+1)}}
F_n \Bigl( \bigl(\frac{\lambda_{2p}a^6}{p} \bigr)^{\frac{1}{2(p+1)}} \phi \Bigr).
\end{eqnarray}
Here, $\epsilon_n$ is independent of $a$. Since the eigenfunctions constitute a basis denoted by an infinite vector $\vec{\Phi}^{\rm T} (\phi, a) = (\Phi_0, \Phi_1, \cdots)$, the rate of change of the eigenfunctions is given by a coupling matrix
\begin{eqnarray}
\frac{\partial}{\partial a} \vec{\Phi} (\phi, a) = \Omega (a) \vec{\Phi} (\phi, a), \label{con tran}
\end{eqnarray}
where $\Omega_{ln} (a)$ is a matrix inversely proportional to $a$ as
\begin{eqnarray}
\Omega_{ln} (a) = \frac{3}{4(p+1) a} \bigl( \epsilon_l - \epsilon_n \bigr) \int d \zeta F_l (\zeta) F_n (\zeta) \zeta^2. \label{coup mat}
\end{eqnarray}
The meaning of Eq. (\ref{con tran}) is continuous transitions among the eigenfunctions at each moment of intrinsic time $a$.

For the Cauchy problem, expand the wave function by the eigenfunctions of the inflaton Hamiltonian, $\Psi (a, \phi) = \vec{\Phi}^{\rm T} (\phi, a) \cdot \vec{\Psi} (a)$, and find the two-component wave function\cite{kim92,kim-page92}
\begin{eqnarray}
\left(
  \begin{array}{c}
    \Psi (a, \phi) \\
    \frac{\partial  \Psi (a, \phi)}{\partial a} \\
  \end{array}
\right) = \left(
            \begin{array}{cc}
             \vec{\Phi}^{\rm T}   & 0 \\
              0 & \vec{\Phi}^{\rm T} \\
            \end{array}
          \right)
 {\rm T} \exp \left[ \int_{a_0}^a \left(
            \begin{array}{cc}
              \Omega & I \\
              V_{\rm G} I - \frac{E}{a'^2} & \Omega \\
            \end{array}
          \right) da' \right]
          \left(
            \begin{array}{c}
             \vec{\Psi} (a_0) \\
              \frac{d \vec{\Psi} (a_0)}{da_0} \\
            \end{array}
          \right), \label{cauchy}
\end{eqnarray}
where $E(a)= (E_0, E_1, \cdots)$. The WDW equation has another Cauchy problem of the Feshbach-Villars formulation.\cite{mostafazadeh}
Taking only the off-diagonal components, which is equivalent to neglecting the coupling matrix $\Omega$, the equation for the gravitational part is approximately given by
\begin{eqnarray}
\frac{d^2 \vec{\Psi} (a)}{da^2}  - \Bigl(V_{\rm G} (a) I - \frac{E(a)}{a^2} \Bigr) \vec{\Psi} (a) \approx 0.
\end{eqnarray}
From the view point of observational cosmology, the task of quantum cosmology is to construct the present Cauchy data based on observations and to evolve back to the early universe to understand the origin of the universe. Note that $\Omega$ diverges more rapidly than $E(a)/a^2$ for $p < 5$ as the universe approaches the Big Bang singularity, $a \approx 0$. For instance, a massive scalar field model with $p=1$ and $\lambda_2 = m^2$ has the harmonic wave functions and the coupling matrix
\begin{eqnarray}
E_n (a) = m a^3 (2n+1), \quad \Omega_{ln} (a) = \frac{3}{4a} \Bigl(\sqrt{l(l-1)} \delta_{l-n,2} - \sqrt{n(n-1)} \delta_{n-l,2} \Bigr),
\end{eqnarray}
and the time-ordered integral is thus approximated by
\begin{eqnarray}
{\rm T} \exp \left[ \int_{a_0}^a \left(
            \begin{array}{cc}
              \Omega (a') & 0 \\
             0 & \Omega (a') \\
            \end{array}
          \right) da' \right] =  \left(
            \begin{array}{cc}
             e^{ \ln (a/a_0) a \Omega (a)} & 0 \\
             0 & e^{ \ln (a/a_0) a \Omega (a)} \\
            \end{array}
          \right).
\end{eqnarray}
Therefore, the wave function experiences an infinite number of transitions among different harmonic functions or oscillations near the singularity, which may lead to a chaotic behavior. Further, the probability for the wave function near the singularity is almost invariant, $|\Psi (a, \phi)|^2 \approx |\Psi (a_0, \phi)|^2$, as afar as the variation of $\vec{\Psi} (a)$ is finite.\cite{kim13}

\section{Semiclassical and Classical Cosmology} \label{semi-class cos}

The wave function peaked around a wave packet allows the de Broglie-Bohm pilot-wave theory. We assume that the WDW equation takes a general form
\begin{eqnarray}
\Bigl[- \frac{\hbar^2}{2M} \nabla^2 - M V_{\rm G} (h_a) + \hat{H}_{\rm M} (\phi, -i \frac{\delta}{\delta \phi} ; h_a) \Bigr] \Psi (h_a, \phi) = 0,
\end{eqnarray}
where $h_a$ denotes the superspace metric $h_{ij}$ and $M = m_P^2$ is the Planck mass squared. The de Broglie-Bohm pilot theory describes the quantum theory in an equivalent way that the oscillating wave function forms a wave packet around a trajectory prescribed by the Hamilton-Jacobi equation with a quantum correction and another equation for the conservation of probability.\cite{licata-fiscaletti} The standard cosmology is the Friedmann equation together with the principle of homogeneity and isotropy of the universe, which has been confirmed precisely by CMB and other observational data. To embody quantum cosmology into an observational cosmology, as shown in Fig. \ref{QSC}, it is necessary to obtain  the semiclassical cosmology and then the classical cosmology with quantum corrections included.

The stratagem toward Fig. \ref{QSC} is first to apply the Born-Oppenheimer idea, which separates a slow moving massive particle $M$ (gravity) from a fast moving light particle (inflaton), and then  to expand quantum state for fast moving variable by a certain basis\cite{kim95,kim96,BFV,kim97}
\begin{eqnarray}
\vert \Psi (h_a, \phi) \rangle = \Psi (h_a) \vert \Phi (\phi, h_a) \rangle,
\end{eqnarray}
in which
\begin{eqnarray}
\vert \Phi (\phi, h_a) \rangle = \vec{C}^T (h_a) \cdot \vec{\Phi} (\phi, h_a).
\end{eqnarray}
The basis $\vec{\Phi}^T (\phi, h_a) = (\vert \Phi_0  \rangle, \vert \Phi_1  \rangle, \cdots)$, which is not necessarily the instantaneous eigenfunctions of $\hat{H}_{\rm M}$, will be chosen to make the semiclassical and classical cosmology as simple as possible. We then apply the de Broglie-Bohm pilot-theory to the gravity part only
\begin{eqnarray}
\Psi(h_a) = F (h_a) e^{i \frac{S(h_a)}{\hbar}}.
\end{eqnarray}
Now, in a semiclassical regime, the WDW equation is equivalent to set of equations\cite{kim97}
\begin{eqnarray}
\frac{1}{2M} \bigl( \nabla S \bigr)^2 - M V_{\rm G} (h_a) + H_{nn} (h_a) - \frac{\hbar^2}{2M} \frac{\nabla^2 F}{F} - \frac{\hbar^2}{M} {\rm Re} \bigl( Q_{nn} \bigr) = 0, \label{quan ein eq} \\
\frac{1}{2} \nabla^2 S + \frac{\nabla F}{F} \cdot \nabla S + {\rm Im} \bigl( Q_{nn} \bigr) = 0, \label{con eq}
\end{eqnarray}
where $H_{nk} (h_a)$ is the expectation value of the inflaton, $\vec{A}_{nk} (h_a)$ is the induced gauge potential due to the parametric interaction, $(\hbar^2/2M) (\nabla^2 F/F)$ is the quantum potential of Bohm and $Q_{nn}$ is the quantum back reaction of matter:
\begin{eqnarray}
H_{nk} (h_a) &=& \langle \Phi_n (\phi, h_a) \vert \hat{H}_{\rm M} \vert \Phi_k (\phi, h_a) \rangle, \nonumber\\
\vec{A}_{nk} (h_a) &=& i \langle \Phi_n (\phi, h_a) \vert \nabla\vert \Phi_k (\phi, h_a) \rangle, \nonumber\\
Q_{nn} (h_a) &=& \frac{\nabla F}{F} \cdot \Bigl( \frac{\nabla C_n}{C_n} - i \sum_{k} \vec{A}_{nk} \frac{C_k}{C_n} \Bigr).
\end{eqnarray}
The advancement of the de Broglie-Bohm quantum theory in Ref. \refcite{kim97} is that the continuity equation (\ref{con eq}) may be integrated along the semiclassical trajectory and provide another quantum back reaction to the semiclassical Einstein equation (\ref{quan ein eq}).

To complete the transition from the quantum cosmology (\ref{wdw eq}) to the semiclassical cosmology, we introduce a cosmological time as the directional derivative along the semiclassical trajectory in the extended superspace
\begin{eqnarray}
\frac{\partial}{\partial \tau} = - \frac{1}{Ma} \frac{\partial S(a)}{\partial a} \frac{\partial}{\partial a}.
\end{eqnarray}
The cosmological time is equivalent to solving $\partial a(\tau)/\partial \tau = - (1/Ma) (\partial S/ \partial a)$. Then the Heisenberg matrix
equation takes the form\cite{kim97}
\begin{eqnarray}
i \hbar \frac{\partial C_n}{\partial \tau} = \sum_{k} \Bigl[ \bigl( H_{nk} - H_{nn} \delta_{nk} \bigr) - \hbar B_{nk}  - \frac{\hbar^2}{2Ma} \bigl( D_{nk} - D_{nn} \delta_{nk} \bigr) \Bigr] C_k, \label{heis eq}
\end{eqnarray}
where $B_{nk}$ is the gauge potential $\vec{A}_{nk}$ measured along the $\tau$-flow and $D_{nk}$ is given by
\begin{eqnarray}
B_{nk} (a(\tau)) &=& i \langle \Phi_n \vert \frac{\partial}{\partial \tau} \vert \Phi_k \rangle, \nonumber\\
D_{nk} (a(\tau)) &=& - \frac{1}{\dot{a}^2} \Bigl[\Bigl(\frac{\partial^2}{\partial \tau^2} - \frac{\ddot{a}}{\dot{a}} \frac{\partial}{\partial \tau}\Bigr)\delta_{nk}  - 2 i B_{nk} \frac{\partial}{\partial \tau} + \langle \Phi_n \vert \frac{\partial^2}{\partial \tau^2} - \frac{\ddot{a}}{\dot{a}} \frac{\partial}{\partial \tau} \vert \Phi_k \rangle \Bigr].
\end{eqnarray}
Here and hereafter, the dots denote the derivatives with respect to the cosmological time $\tau$. We may use the freedom to select the basis such that $H_{nk} = \hbar B_{nk}$ for $n \neq k$.\cite{kim96} Then, the Heisenberg equation (\ref{heis eq}) is the solution of the $\tau$-dependent Schr\"{o}dinger equation and
a correction of order of $\hbar/M$, which comes from the relativistic theory. Now, we can show thereby that the chaotic inflation model necessarily contains (higher) curvature terms
\begin{eqnarray}
\Bigl(\frac{\dot{a}}{a} \Bigr)^2 + \frac{k}{a^2} - \Lambda = \frac{8 \pi}{3m_P^2 a^3} \bigl( H_{nn} + \delta \rho_{nn} \bigr), \label{qc frw eq}
\end{eqnarray}
where the quantum correction to the energy density is given by
\begin{eqnarray}
\delta \rho_{nn} =  - \frac{4 \pi \hbar^2}{3m_P^2 a \dot{a}} U_{nn} {\rm Re} \bigl(R_{nn} \bigr) + \frac{2 \pi \hbar^2}{3m_P^2 a } \Bigl( U_{nn}^2 + \frac{1}{\dot{a}} \dot{U}_{nn} \Bigr), \label{qc cor}
\end{eqnarray}
where
\begin{eqnarray}
R_{nn} &=& \frac{\dot{C}_n}{C_n} - i \sum_{k} B_{nk} \frac{C_k}{C_n}, \nonumber\\
U_{nn} &=& - \frac{1}{2} \frac{\dot{a}^2 + a \ddot{a}}{a \dot{a}^2 + \frac{4 \pi \hbar}{3m_P^2} {\rm Im} \bigl(R_{nn} \bigr)}.
\end{eqnarray}

Finally, the chaotic model with a massive scalar with $m$ has $R_{nn}^{(0)} = 0$ and $U_{nn}^{(0)} = -(1/2) (1/a+ \ddot{a}/\dot{a}^2)$ and the expectation value of the inflaton is given by\cite{kim97}
\begin{eqnarray}
H_{nn} = \hbar a^3 \Bigl(n+ \frac{1}{2} \Bigr) \bigl(\dot{\varphi}^* \dot{\varphi} + m^2 \varphi^* \varphi \bigr),
\end{eqnarray}
which obeys the classical equation of motion
\begin{eqnarray}
\ddot{\varphi} + 3 \frac{\dot{a}}{a} \dot{\varphi} + m^2 \varphi = 0.
\end{eqnarray}
The gauge potential reads
\begin{eqnarray}
B_{nk} = \frac{n}{n+ \frac{1}{2}} H_{nn} \delta_{nk} + f \sqrt{(n+1)(n+2)} \delta_{n, k-2} + f^* \sqrt{(k+1)(k+2)} \delta_{n, k+2},
\end{eqnarray}
where $f = - (\hbar a^3 /2) (\dot{\varphi}^2 + m^2 \varphi^2 )$. The solution of the Heisenberg equation (\ref{heis eq}) can be used to find systematically the quantum correction (\ref{qc cor}) and thus to the Friedmann equation (\ref{qc frw eq}).

\section{Gauge Invariant Quantum Cosmology}

Mukhanov and Sasaki obtained the gauge invariant formulation of density perturbations. Then a question can be raised whether one may find the Hamiltonian for scalar perturbations of metric and a massive scalar field, which is gauge invariant and leads the semiclassical equation for observational data. In fact, such a Hamiltonian was found\cite{pinho-pinto,PSS,CFMO,CMM}
\begin{eqnarray}
H = \bar{N}_0 \Bigl[H_0 + \sum_{\vec{n}, \pm} \breve{H}_2^{\vec{n}, \pm}  \Bigr] + \sum_{\vec{n}, \pm} \breve{G}_2^{\vec{n}, \pm} \breve{H}_1^{\vec{n}, \pm}
+ \sum_{\vec{n}, \pm} \breve{K}_{\vec{n}, \pm} \tilde{H}_1^{\vec{n}, \pm}, \label{MS ham}
\end{eqnarray}
where $H_0$ is the unperturbed Hamiltonian for the FRW universe, and $\breve{H}_2^{\vec{n}, \pm}$ is the quadratic Hamiltonian of scalar, vector and tensor perturbations for the inhomogeneities as well as the massive scalar field, and $\breve{H}_1^{\vec{n}, \pm}$ and $\tilde{H}_1^{\vec{n}, \pm}$  are inhomogeneous linear perturbations. Extending Sec. \ref{semi-class cos} to quadratic perturbations, one may show that the semiclassical cosmology from the WDW equation (\ref{MS ham}) provides the master equation for the power spectrum of primordial scalar (vector and tensor) perturbations perturbations. The semiclassical cosmology with (higher) curvatures via the de Broglie-Bohm pilot theory and the Born-Oppenheimer idea may resurrect the chaotic inflation model with a massive scalar field. The detailed work in this direction will be addressed in a future publication.

\section{Conclusion}

In this paper we have studied the semiclassical gravity for the chaotic inflation model with a power-law greater than one. In classical gravity theory the chaotic inflation model with a convex power-law is highly likely to be excluded by current observational CMB data, such Planck and WMAP. Starobinsky model with a higher curvature term, however, seems to be a viable model. Higher curvature terms have a quantum origin due to fluctuations of a spacetime and/or a matter field in the curved spacetime. It is thus physically legitimate to investigate the chaotic inflation model in the framework of quantum cosmology since the quantum cosmological model with a chaotic inflaton necessarily involves quantum gravity effects due to spacetime and inflaton's fluctuations.

In order to compare the predictions of the quantum cosmology with current observational data, the semiclassical cosmology should emerge from the quantum cosmology for the FRW cosmology minimally coupled to a chaotic inflaton, in particular, a massive inflaton. In fact, the de Broglie-Bohm pilot-theory together with the Born-Oppenheimer idea of separating the Planck mass scale from the inflaton mass scale leads to the semiclassical cosmology, in which both quantum corrections to the classical gravity as well as the matter field. It turns out that the semiclassical gravity equation indeed contains higher curvature terms for the FRW geometry. Further, the gauge invariant quantum cosmology using Mukhanov-Sasaki Hamiltonian with a massive scalar field may yield the semiclassical chaotic inflation model, which may be easily compared with observational  data. It would be interesting to study numerically the semiclassical gravity for the FRW universe with a massive scalar field and to see whether the quantum cosmology can resurrect the chaotic inflation model.

An alternative way to test the predictions of quantum cosmology  is to simulate the evolution of the universe, in particular, the quantum effects of an expanding universe using laboratory experiments. It has been suggested that a static ion trap may simulate the quantum effects of expanding universe.\cite{MOM} It has also been observed that the quantum cosmology for the FRW universe minimally coupled to a massive scalar field is equivalent to a spinless charged particle in a homogeneous time-dependent magnetic field along a fixed direction.\cite{kim14b} The infinite oscillations near the singularity can be simulated by time-dependent magnetic fields of over-critical strength.

\section*{Acknowledgments}
This research was supported by Basic Science Research Program through the National Research Foundation of Korea (NRF) funded by the Ministry of Education (15B15770630).



\begin{thebibliography}{0}    

\bibitem{LIGO col} B.~P.~Abbot et al (LIGO Scientific Collaboration and Virgo Collaboration), {\it Phys.\ Rev.\ Lett.}\ {\bf 116}, 061102 (2016).

\bibitem{hawking-penrose} S.~W.~Hawking and R.~Penrose, {\it Proc.\ Roy.\ Soc.\ Lond.\ A} {\bf 314}, 529 (1970).

\bibitem{BGV} A.~Borde, A.~H.~Guth, and A.~Vilenkin, {\it Phys.\ Rev.\ Lett.}\ {\bf 90}, 151301 (2003).

\bibitem{dewitt67} B.~S.DeWitt, {\it Phys.\ Rev.}\ {\bf 160}, 1113 (1967).

\bibitem{hartle-hawking} J.~B.~Hartle and S.~W.~Haking, {\it Phys.\ Rev.\ D} {\bf 28}, 2960 (1983).

\bibitem{hawking84} S.~W.~Hawking, {\it Nucl.\ Phys.\ B}\ {\bf 239}, 257 (1984).

\bibitem{coule} D.~H.~Coule, {\it Class. Quantum Grav.} {\bf 22},  R125 (2005).

\bibitem{page06} D.~N.~Page, in {\it The Future of Theoretical Physics and Cosmology} edited by G.~W.~Gibbons, E.~P.~Shellard and S.~J.~Rankin (Cambridge University Press, Cambridge, 2003) [hep-th/0610121].

\bibitem{hawking-page} S.~W.~Hawking and D.~N.~Page, {\it Nucl.\ Phys.\ B}\ {\bf 264}, 185 (1986).

\bibitem{halliwell-hawking} J.~J.~Halliwell and S.~W.~Hawking, {\it Phys.\ Rev.\ D} {\bf 31}, 1777 (1985).

\bibitem{hawking85} S.~W.~Hawking {\it Phys.\ Rev.\ D}\ {\bf 32}, 2489 (1985).

\bibitem{page85} D.~N.~Page, {\it Phys.\ Rev.\ D}\ {\bf 32}, 2496 (1985).

\bibitem{HLL} S.~W.~Hawking, R.~Laflamme and G.~W.~Lyons, {\it Phys.\ Rev.\ D}\ {\bf 47}, 5342 (1993).

\bibitem{kim-kim94} S.~P.~Kim and S-W.~Kim, {\it Phys.\ Rev.\ D}\ {\bf 49}, R1679 (1994).

\bibitem{kim-kim95} S.~P.~Kim and S-W.~Kim, {\it Phys.\ Rev.\ D}\ {\bf 51}, 4254 (1995).

\bibitem{kiefer87} C.~Kiefer, {\it Class.\ Quantum\ Grav.}\ {\bf 4}, 1369 (1987).

\bibitem{kim92} S.~P.~Kim, {\it Phys.\ Rev.\ D}\ {\bf 46}, 3403 (1992).

\bibitem{starobinsky} A.~Starobinsky, Present status of inflation and future perspectives, talk at International Conference on Gravitation and Cosmology, KITPC, Beijing, May 4-8, 2015.

\bibitem{Planck2013} G. Hinshaw et al, {\it Astrophys.\ J.\ Suppl.}\ {\bf 208}, 19 (2013).

\bibitem{starobinsky80} A.~Starobinsky, {\it Phys.\ Lett.}\ {\bf 91}, 99 (1980).

\bibitem{whitt84} B.~Whitt, {\it Phys.\ Lett.}\ {\bf 145}, 176 (1984).

\bibitem{maeda88} K.~Maeda, {\it Phys.\ Rev.\ D}\ {\bf 37}, 858 (1988).

\bibitem{kim13} S.~P.~Kim, {\it The Universe} {\bf 1}, 11 (2013) [arXiv:1304.7439].

\bibitem{kim14} S.~P.~Kim, {\it Nucl.\ Phys. B\ Proc. Suppl.} {\bf 246}, 68 (2014).

\bibitem{kim95} S.~P.~Kim, {\it Phys.\ Rev.\ D}\ {\bf 52}, 3382 (1995).

\bibitem{kim96} S.~P.~Kim, {\it Class.\ Quantum\ Grav.}\ {\bf 13}, 1377 (1996).

\bibitem{BFV} C.~Bertoni,F.~Finelli, and G.~Venturi, {\it Class.\ Quantum\ Grav.}\ {\bf 13}, 2375 (1996).

\bibitem{kim97} S.~P.~Kim, {\it Phys.\ Rev.\ D}\ {\bf 55}, 7511 (1997).

\bibitem{kiefer-kramer} C.~Kiefer and M.~Kr\"{a}mer, {\it Phys.\ Rev.\ Lett.}\ {\bf 108}, 021301 (2012).

\bibitem{kk-essay} C.~Kiefer and M.~Kr\"{a}mer, {\it Int.\ J.\ Mod.\ Phys.\ D}\ {\bf 21}, 1241001 (2012).

\bibitem{pinho-pinto} E.~ J.~ C.~Pinho and N.~Pinto-Neto, {\it Phys.\ Rev.\ D}\ {\bf 76}, 023506 (2007).

\bibitem{PSS} N.~Pinto-Neto, G.~Santos, and W.~Struyve, {\it Phys.\ Rev.\ D}\ {\bf 85}, 083506 (2012).

\bibitem{CFMO} L.~Castell\'{o} Gomar, M.~Fern\'{a}ndez-M\'{e}ndez, G.~ A.~ Mena Marug\'{a}, and J.~Olmedo, {\it Phys.\ Rev.\ D}\
{\bf 90}, 064015 (2014).

\bibitem{CMM} L.~Castell\'{o} Gomara, M.~ Mart\'{i}n-Benitob and G.~ A.~Mena Marug\'{a}n, {\it JCAP} {\bf 06} (2015) 045.

\bibitem{kim-page92} S.~P.~Kim and D.~N.~Page, {\it Phys.\ Rev.\ D}\ {\bf 45}, R3296 (1992).

\bibitem{mostafazadeh} A.~Mostafazadeh, {\it J.\ Math.\ Phys.}\ {\bf 39}, 4499 (1998).

\bibitem{licata-fiscaletti} I.~Licata and D.~Fiscaletti, {\it Quantum Potential: Physics, Geometry and Algebra} (Springer-Verlag, New York, 2014).

\bibitem{MOM} N.~C.~Menicucci, S.~J.~Olson and G.~J.~ Milburn, {\it New J. Phys.} {\bf 12}, 095019 (2010).

\bibitem{kim14b} S.~P.~Kim, {\it Ann. Phys.} {\bf 351}, 54 (2014).

\end{thebibliography}
\end{document}